\documentclass[aip,a4paper,apl,preprint,showkeys,showpacs]{revtex4-1} 
\usepackage{here}
\usepackage{graphicx}
\usepackage{amsmath}
\usepackage{geometry}
\usepackage{amssymb}
\usepackage{bm}
\usepackage{times}
\usepackage{fancyhdr}
\usepackage{color}
\usepackage{multirow}
\usepackage{here} 

\begin{document}

\title{ Magnetic anisotropy of $\bm{L1_0}$-ordered FePt thin films studied by  Fe and Pt $\bm{L_{2,3}}$-edges x-ray magnetic circular dichroism}
\date{\today}

\begin{abstract}
The strong perpendicular magnetic anisotropy of $L{\rm1_0}$-ordered FePt has been the subject of extensive studies for a long time. However, it is not known which element, Fe or Pt, mainly contributes to the magnetic anisotropy energy (MAE). We have investigated  the anisotropy of the orbital magnetic moments of Fe 3$d$ and Pt 5$d$ electrons in $L{\rm1_0}$-ordered FePt thin films by Fe and Pt $L_{2,3}$-edge x-ray magnetic circular dichroism (XMCD) measurements for samples with various degrees of long-range chemical order $S$. Fe $L_{2,3}$-edge XMCD showed that the orbital magnetic moment was larger when the magnetic field was applied perpendicular to the film than parallel to it, and that the anisotropy of the orbital magnetic moment increased with $S$. Pt $L_{2,3}$-edge XMCD also showed that the orbital magnetic moment was smaller when the magnetic field was applied perpendicular to the film than parallel to it, opposite to the Fe $L_{2,3}$-edge XMCD results although the anisotropy of the orbital magnetic moment increases with $S$ like the Fe edge. These results are qualitatively consistent with the first-principles calculation by Solovyev {\it et\ al.} [Phys. Rev. B {\bf 52}, 13419 (1995).], which also predicts the dominant contributions of Pt 5$d$ to the magnetic anisotropy energy rather than Fe 3$d$ due to the strong spin-orbit coupling and the small spin splitting of the Pt 5$d$ bands in $L{\rm1_0}$-ordered FePt.
\end{abstract}

\author{\mbox{K. Ikeda}}
\affiliation{Department of Physics, University of Tokyo, Hongo 113-0033, Japan}
\email{k-ikeda@wyvern.phys.s.u-tokyo.ac.jp}

\author{T. Seki}
\affiliation{Institute for Materials Research, Tohoku University, Sendai 980-8577, Japan}

\author{G. Shibata}
\affiliation{Department of Physics, University of Tokyo, Hongo 113-0033, Japan}

\author{T. Kadono}
\affiliation{Department of Physics, University of Tokyo, Hongo 113-0033, Japan}

\author{\mbox{K. Ishigami}}
\affiliation{Department of Physics, University of Tokyo, Hongo 113-0033, Japan}

\author{\mbox{Y. Takahashi}}
\affiliation{Department of Physics, University of Tokyo, Hongo 113-0033, Japan}

\author{\mbox{M. Horio}}
\affiliation{Department of Physics, University of Tokyo, Hongo 113-0033, Japan}

\author{\mbox{S. Sakamoto}}
\affiliation{Department of Physics, University of Tokyo, Hongo 113-0033, Japan}

\author{\mbox{Y. Nonaka}}
\affiliation{Department of Physics, University of Tokyo, Hongo 113-0033, Japan}

\author{\mbox{M. Sakamaki}}
\affiliation{Institute of Materials Structure Science, KEK, Tsukuba 305-0801, Japan}

\author{\mbox{K. Amemiya}}
\affiliation{Institute of Materials Structure Science, KEK, Tsukuba 305-0801, Japan}

\author{\mbox{N. Kawamura}}
\affiliation{JASRI, 1-1-1 Kouto, Sayo, Hyogo 679-5198, Japan}

\author{\mbox{M. Suzuki}}
\affiliation{JASRI, 1-1-1 Kouto, Sayo, Hyogo 679-5198, Japan}

\author{\mbox{K. Takanashi}}
\affiliation{Institute for Materials Research, Tohoku University, Sendai 980-8577, Japan}

\author{\mbox{A. Fujimori}}
\affiliation{Department of Physics, University of Tokyo, Hongo 113-0033, Japan}

\maketitle


\graphicspath{{fig/}}
In recent years, the amount of data we deal with has been increasing and the information technology has to adapt to the era of big data. To this end, the enhancement of the recording density in magnetic recording media has been highly desired. In the magnetic recording media like hard disk drives, larger data capacity can be achieved by using materials with perpendicular magnetization and by decreasing the bit size, but it also leads to the reduction of thermal stability. To maintain the thermal stability, it is important to employ magnetic materials with larger magnetic anisotropy energy (MAE). $L{\rm1_0}$-ordered FePt is one of such materials that have the largest MAE with the perpendicular easy magnetization axis, and is a candidate to achieve higher density of magnetic recording media. The microscopic origin of the strong perpendicular magnetic anisotropy of $L{\rm1_0}$-ordered FePt, however, has not been fully understood yet despite its importance and extensive studies.

 In order to reveal the origin of the MAE of $L{\rm1_0}$-ordered FePt, magnetism at the Fe sites and the Pt sites need to be studied by an element specific technique. Especially, the orbital magnetic moments of Fe and Pt have to be investigated separately since it has been well known that the orbital magnetic moment plays an important role in magnetic anisotropy\ \cite{Bruno_expansion_Laan, Bruno_MAE_3dmonolayer}. 
X-ray magnetic circular dichroism (XMCD) at the $L_{\rm 2,3}$ edges is an ideal tool to investigate the magnetic properties of $d$ electrons in transition-metal alloys in an element specific way. In the present study, we have studied the spin ($m_{\text{spin}}$) and orbital magnetic moments ($m_{\text{orb}}$) of Fe 3$d$ and Pt 5$d$ electrons and their anisotropy using Fe and Pt $L$-edge XMCD and discuss the microscopic origin of the MAE of $L{\rm1_0}$-ordered FePt thin films.

FePt thin films were grown on MgO (100) substrates by the ultrahigh vacuum dc-sputtering method with Fe and Pt targets. The stacked structure of the sample was MgO substrate/Fe (1 nm)/Au (30 nm)/FePt (20 nm)/Au (2 nm). The 2nm-thick top Au layer was deposited as a cap layer and the bottom Au layer of 30 nm thickness beneath the FePt layer was inserted as a buffer layer to reduce the effect of the lattice mismatch between FePt and the MgO substrate.
The deposition temperature ($T_S$) and the annealing temperature ($T_A$) ranged from room temperature to 600 ${}^\circ\mathrm{C}$. 
Table\ \ref{table:PFBL16_energy} summarizes the relationship between sample-preparation conditions, the degree of long-range chemical order $S$ estimated using x-ray diffraction\ \cite{FePt_fund_Seki}, and the MAE. Both $S$ and MAE increase with $T_S$ and $T_A$, consistent with the previous study\ \cite{FePt_fund_Seki}. The MAE of the sample with $S=0$ shows a small negative value, which is due to errors in the estimation of the MAE from the {\it M-H} curve (Fig. S.1).
\begingroup
\squeezetable
\begin{table}[t]
\begin{center}
\begin{ruledtabular}
\caption{
Relationship between the sample-preparation conditions, the degree of long-range chemical order $S$, and the magnetic anisotropy energy (MAE) of samples studied in the present work. The MAE has been estimated from the {\it M-H} curves (Fig. S1). The shape anisotropy energy has been subtracted from the MAE.}
\label{table:PFBL16_energy}
\vspace*{0.3cm}
\begin{tabular}{cccc}
 \shortstack{Deposition \\temperature (${}^\circ\mathrm{C}$)} & \shortstack{Annealing \\temperature (${}^\circ\mathrm{C}$)} & \shortstack{Degree of long-range \\chemical order parameter $S$} & \shortstack{MAE\\ ($\rm{MJ}/m^3$)}	\\
\hline 
300 & 600 & 0.7 $\pm$ 0.1 & 4.6	\\

300 & 500 & 0.5 $\pm$ 0.1 & 1.6	\\

300 & -- & 0.4 $\pm$ 0.1 & 1.3	\\

RT & -- & 0 & -0.37	\\
\end{tabular}
\end{ruledtabular}
\end{center}
\end{table}
%


Fe $L$-edge XMCD measurements were performed at the undulator beamline BL-16A1 of Photon Factory, High Energy Accelerator Research Organization (KEK-PF), and the Pt $L$-edge XMCD measurements at the undulator beamline BL39XU of SPring-8. All the measurements were performed at room temperature. The magnitude of the magnetic field was 5 T at BL-16A1 and 7 T at BL39XU. The magnetic field was applied parallel or anti-parallel to incident x rays. In order to measure the in-plane magnetic moments, the incident angle $\theta$ of x rays measured from the sample surface (as shown in the inset of Fig.\ \ref{fig:Fe_XASXMCD_FePt} and Fig.\ \ref{fig:Pt_XAS_FePt}) was set to 30${}^\circ$ and 4.4${}^\circ$, respectively, for the Fe $L$-edge and Pt $L$-edge XMCD measurements. For the Fe $L$-edge XMCD measurements, four x-ray absorption spectroscopy (XAS) spectra were taken by inverting the external magnetic field and the helicity of x rays independently, and then were averaged. For the Pt $L$-edge XMCD measurements, the helicity switching mode was used. The external magnetic field was inverted and then the obtained two XMCD spectra were averaged. Fe $L$-edge XAS and XMCD were measured in the total electron-yield detection mode and Pt $L$-edge XAS and XMCD were measured in the partial fluorescence-yield detection mode with silicon drift detector.
%
%

Figure\ \ref{fig:Fe_XASXMCD_FePt} shows the XAS and XMCD spectra at the Fe $L_{\rm 2,3}$ edge of {\sl L}1$_{0}$-FePt thin films ($S$ = 0, 0.5, 0.7) under magnetic field $\mu_0H = 5\rm$ T applied parallel (in-plane) and perpendicular to the films (out-of-plane). Each XAS spectrum has been normalized to the height of the $L_{\rm3}$ peak. All the spectra show line shapes similar to that of metallic Fe \cite{Fe_XAS_ref}, and no oxidation features are observed.
 As shown in Fig.\ \ref{fig:Fe_XASXMCD_FePt}(c), the XMCD spectra are different between the two field directions, showing significant anisotropies: the Fe $L_{\rm 3}$-edge XMCD intensity becomes more anisotropic with increasing $S$ while the Fe $L_{\rm 2}$-edge intensity remains nearly isotropic. According to the XMCD sum rules\ \cite{sumrule_Thole}, this behavior indicates the increase of the anisotropy of the orbital magnetic moment with $S$.
\begin{figure}[t]
\includegraphics[width=10cm]{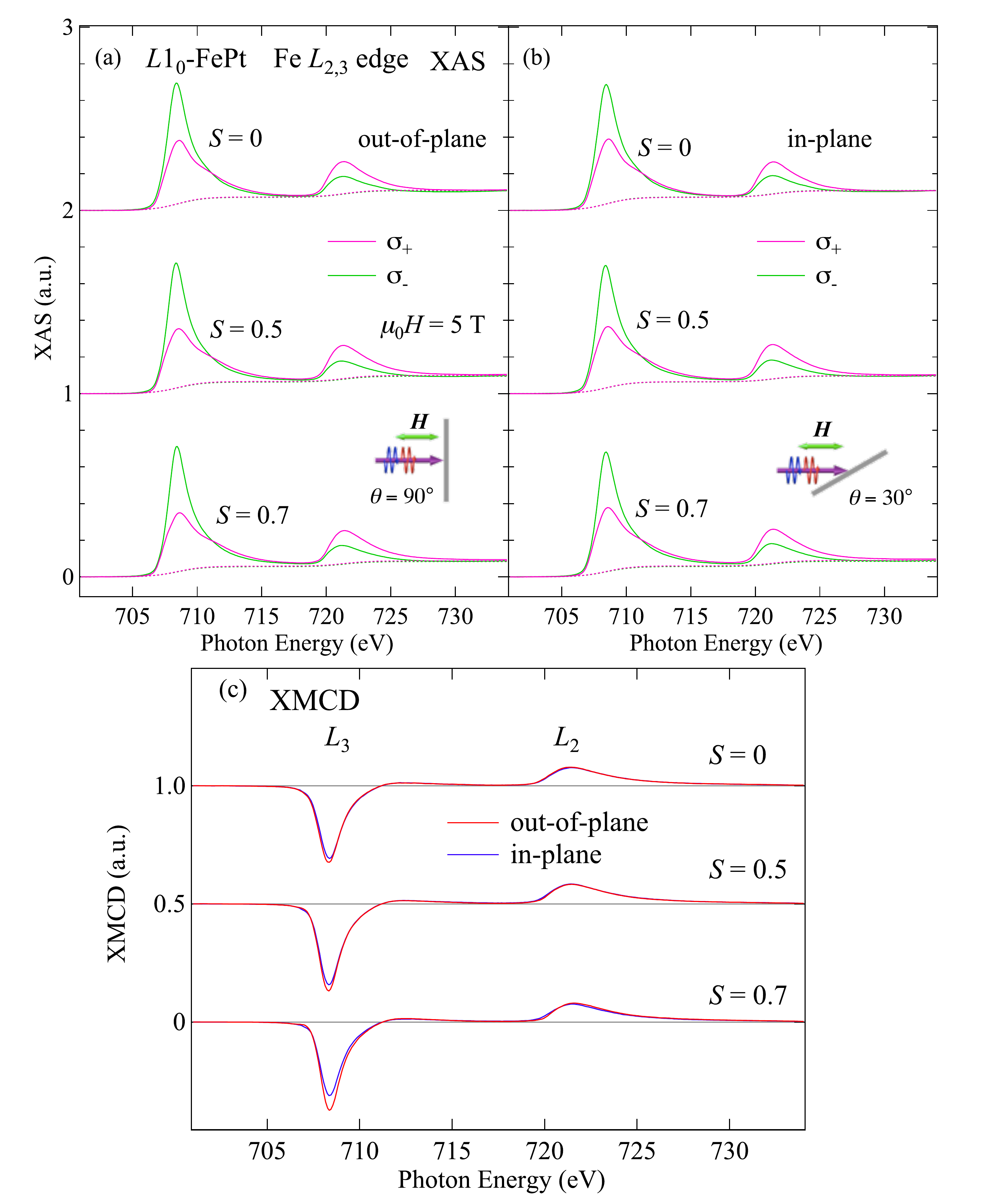}	
\caption{ Fe $L_{\rm2,3}$-edge x-ray absorption spectroscopy (XAS) and x-ray magnetic circular dichroism (XMCD) spectra of {\sl L}1$_{0}$-ordered FePt thin films. (a)(b) XAS spectra for the positive (pink) and negative (green) photon helicities for the out-of-plane (a) and in-plane (b) 
magnetic field directions. $S$ denotes the degrees of long-range chemical order. (c) Comparison of the XMCD spectra between the out-of-plane and in-plane magnetic field directions. All the spectra were taken at room temperature (RT). }
\label{fig:Fe_XASXMCD_FePt}
\end{figure}
%
By applying the XMCD sum rules\ \cite{sumrule_Thole,sumrule_Carra} to the XAS and XMCD spectra, the effective spin ($m^{\text{eff}}_{\text{spin}} = m_{\text{spin}} + 7m_{\text T}$) and orbital magnetic moments have been deduced and plotted in Fig.\ \ref{fig:Fe_moments_FePt}. The hole number  $n_h$ in the sum rules has been assumed to be 3.4 (Ref.\ \onlinecite{Antoniak_FePt_NanoP}). 
The figure shows that the orbital magnetic moment becomes highly anisotropic with increasing $S$ whereas the spin magnetic moments of Fe is nearly isotropic for $S = 0\ \rm{and}\ 0.5 $. 
\begin{figure}[b]
\begin{center}
\includegraphics[width=12cm]{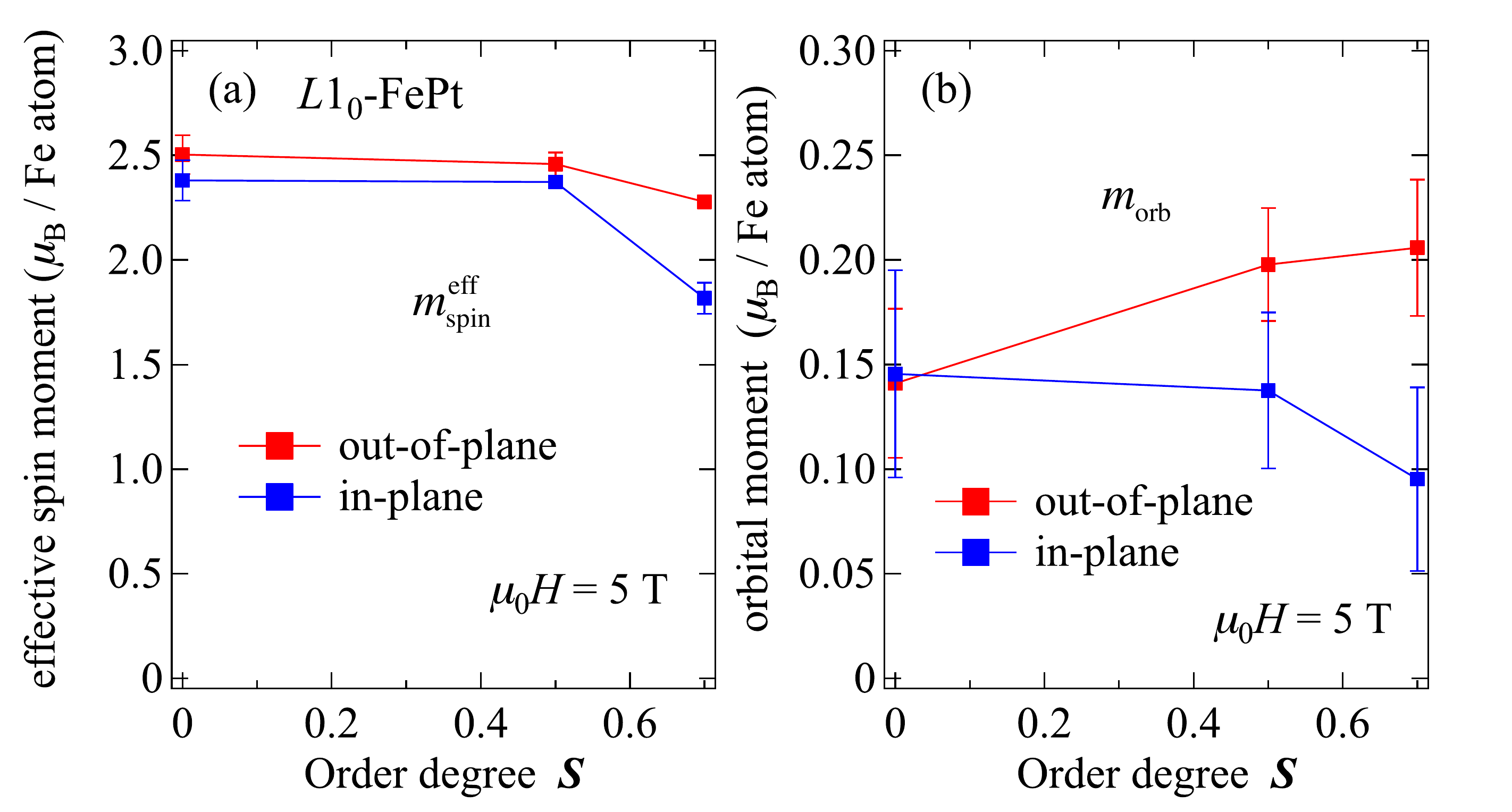}		
\caption{Effective spin magnetic moment $m^{\text{eff}}_{\text{spin}}$ (a) and the orbital magnetic moment $m_{\text{orb}}$ (b)  of Fe in FePt film derived from Fe $L_{\text2,3}$-edge XMCD using the sum rules\ \cite{sumrule_Thole,sumrule_Carra}. The anisotropy of $m^{\text{eff}}_{\text{spin}}$ for the $S=0.7$ sample is due to insufficient magnetic field of $\mu_0H = 5\text{\ T}$ to saturate the in-plane magnetization. Likewise, the anisotropy of $m_{\text{orb}}$ for the same sample is overestimated to a similar extent.}\label{fig:Fe_moments_FePt}
\end{center}
\end{figure}
The spin magnetic moment of the $S=0.7$ sample is supposed to be isotropic, but  the in-plane XMCD spectrum of the $S = 0.7$ sample could not be measured under a magnetically saturated condition because the film with $S = 0.7$ is not magnetically saturated up to about 10 T (see Supplementary Materials Fig.\ S.1). For the same reason, the anisotropy of the the orbital magnetic moment of the $S = 0.7$ sample is overestimated.
%

Figures \ref{fig:Pt_XAS_FePt} shows the Pt $L_{\rm 2,3}$-edge XAS and XMCD spectra of the $L1_{\rm0}$-FePt thin films ($S$ = 0, 0.4, 0.5, 0.7) for out-of-plane and in-plane magnetic fields. Since the Pt $L_2$ and $L_3$ edges are separated by a large energy of $\sim$1.8 keV, the spectra have to be measured and normalized separately, as shown in separate panels (a) and (b) in Fig.\ \ref{fig:Pt_XAS_FePt}. Here, the ratio of the edge jump heights at the $L_3$ and $L_2$ edges have been normalized to 2.22 to 1 (Refs. \onlinecite{FePt_Pt_XMCD_Grange_PRB,Pt_XMCD_Au_whiteL_Bartolome}).
\begin{figure}[t]
\includegraphics[width=10cm]{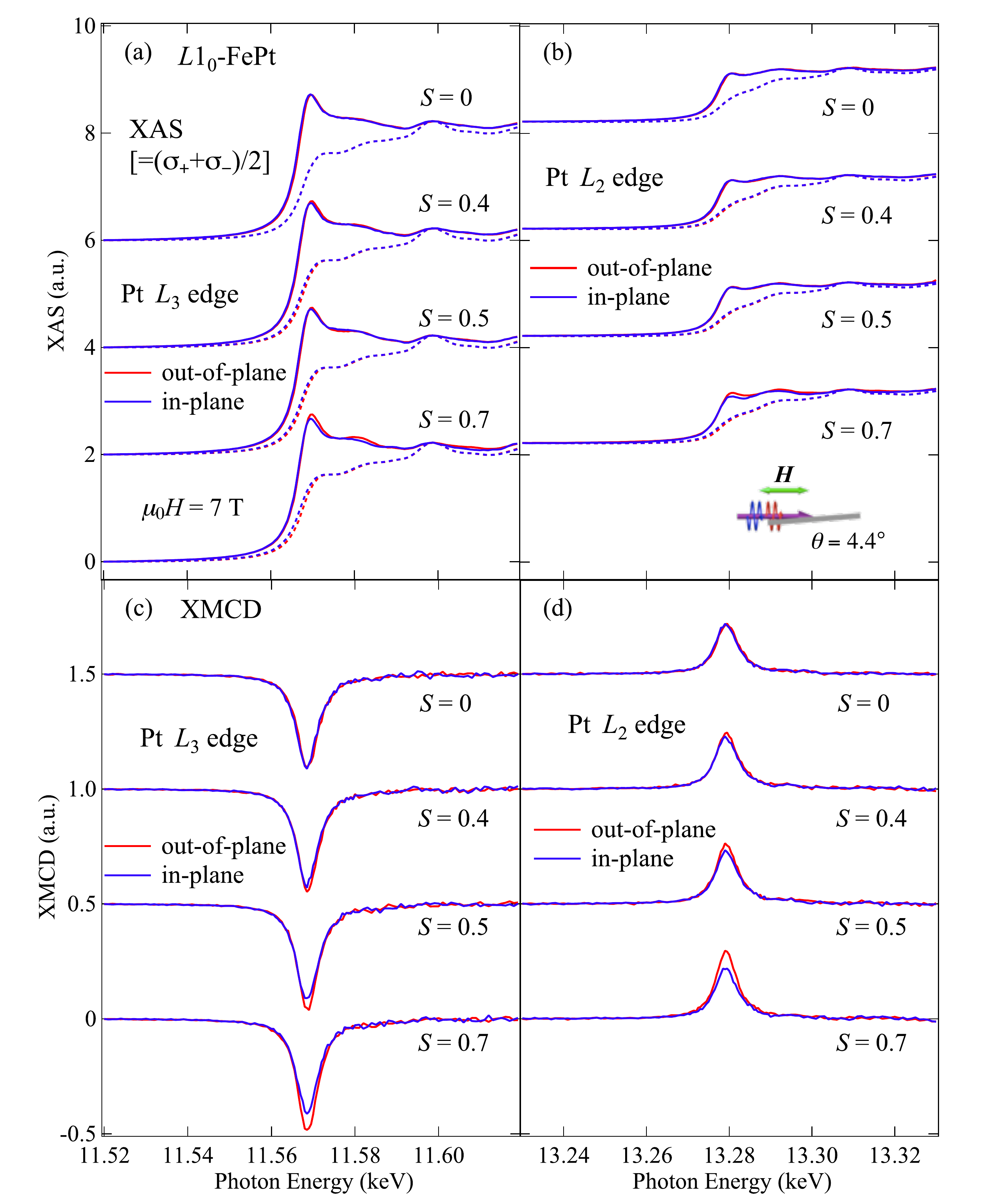}	
\caption{Helicity-averaged XAS spectra of $L1_{\rm0}$-ordered FePt thin films at the  Pt $L_3$ (a) and  Pt $L_2$ edges (b) and the XMCD spectra at the  Pt $L_3$ (c) and  Pt $L_2$ edges (d). The dashed curves are Au XAS spectra.
}
\label{fig:Pt_XAS_FePt}
\end{figure}
As shown in Figs.\ \ref{fig:Pt_XAS_FePt}(c) and \ \ref{fig:Pt_XAS_FePt}(d), the anisotropy of the XMCD intensity increases not only at the $L_{\rm3}$ edge but also at the $L_{\rm2}$ edge with $S$, in contrast to Fe $L_{\text2,3}$-edge XMCD, where only the anisotropy of $L_3$-edge XMCD increases [see Fig.\ \ref{fig:Fe_XASXMCD_FePt}(c)]. The spin and orbital magnetic moments of Pt estimated using the XMCD sum rules are plotted in Fig.\ \ref{fig:Pt_moments_FePt} as functions of $S$.
The intensities of the XAS spectra, which appears in the XMCD sum rules \cite{sumrule_Thole,sumrule_Carra}, 
have been estimated by the similar method as Refs.\ \onlinecite{FePt_Pt_XMCD_Grange_PRB,Pt_XMCD_Au_whiteL_Bartolome,FePt_Pt_XMCD_PRL_Antoniak}. 
First, the XAS spectra of a gold foil 
has been subtracted from the raw XAS spectra of FePt 
so that the background above the main absorption peak vanishes. 
Then, assuming that the intensity of the subtracted spectra $I_\text{diff}$
is proportional to the difference of the hole numbers of 
Pt ($n_h^\text{Pt}$) and Au ($n_h^\text{Au}$), 
we have deduced the XAS intensity of Pt $I_\text{Pt}$ as 
$I_\text{Pt} = \frac{n_h^\text{Pt}}{n_h^\text{Pt}-n_h^\text{Au}}I_\text{diff}$
The hole numbers $n_h$ have been assumed to be 1.73 for Pt and 0.75 for Au \cite{FePt_Pt_XMCD_Grange_PRB,Pt_XMCD_Au_whiteL_Bartolome}.
The magnitudes of the spin and orbital magnetic moments are much smaller than those of Fe, as expected from the fact that Pt is a paramagnetic metal and its ferromagnetic moment is induced through hybridization with  the Fe 3$d$ orbitals. 
More significantly, the sign of the orbital moment anisotropy $\Delta m_{\text{orb}} (= m^{\text{out}}_{\text{orb}} - m^{\text{in}}_{\text{orb}})$ is opposite to that of Fe. 
Note that the in-plane XMCD spectrum of the $S = 0.7$ sample could not be measured under a magnetically saturated condition with $\mu_0H = 7$ T and, therefore, that the weak anisotropy of the spin magnetic moment of that sample would not be intrinsic. For this reason, the anisotropy of the orbital magnetic moment is slightly underestimated. 

\begin{figure}[h]
\begin{center}
\includegraphics[width=12cm]{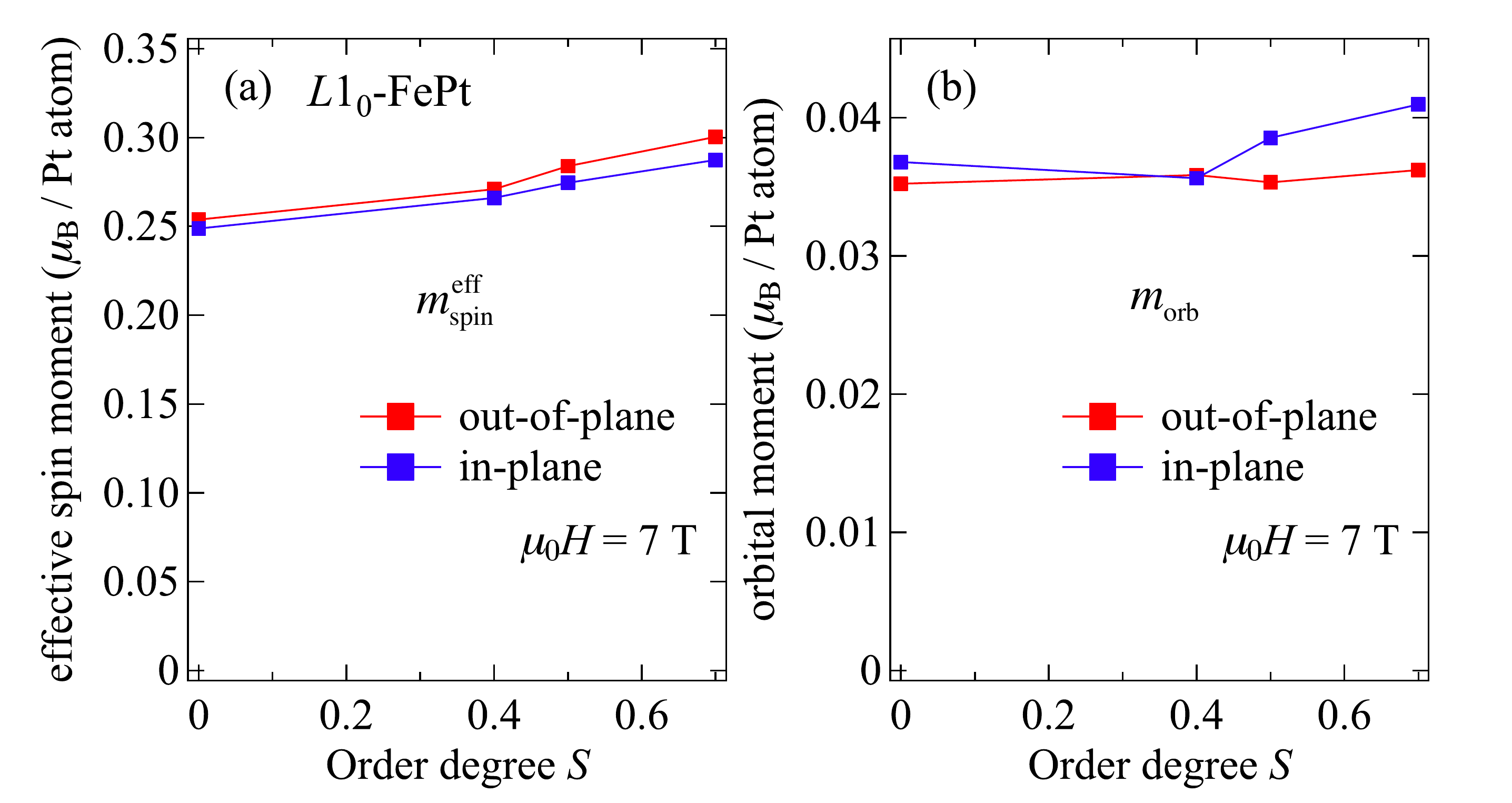} 
\caption{Effective spin magnetic moment $m^{\text{eff}}_{\text{spin}}$ (a) and the orbital magnetic moment $m_{\text{orb}}$ (b) of Pt derived from Pt $L_{\text2,3}$-edge XMCD using the sum rules. The weak anisotropy of $m^{\text{eff}}_{\text{spin}}$ for the $S=0.7$ sample is due to the insufficient magnetic field of $\mu_0H = 7$ T to saturate the in-plane magnetization. Likewise, the anisotropy of $m_{\text{orb}}$ for the same sample is slightly underestimated.  
}
\label{fig:Pt_moments_FePt}
\end{center}
\end{figure}
%
%
%
%
%
%
%
 Here we shall discuss the relationship between the MAE and the observed anisotropy of the orbital magnetic moment. Van der Laan has shown by treating the spin-orbit coupling as a perturbation that the MAE ($\Delta E$) are given by the following equations\ \cite{Bruno_expansion_Laan}.
\begin{eqnarray}
  \Delta E 
   &=& -\frac {\xi}{4\mu_B}\hat{\bm S} \cdot(\langle{\bm L^ {\downarrow}}\rangle^{\text{hard}}-\langle{\bm L^ {\downarrow}}\rangle^{\text{easy}}) \nonumber \\&&+\frac {21}{2} \frac {\xi^2}{\Delta _{\text{ex}}}\hat{\bm S} \cdot(\langle{\bm T}\rangle^{\text{hard}} - \langle{\bm T}\rangle^{\text{easy}} ), 
\label{eq:Etot_laan}
\end{eqnarray}
where $\xi$ is the spin-orbit coupling constant, $\langle{\bm L}\rangle$ is the orbital magnetic moment, $\langle{\bm T}\rangle$ is the magnetic dipole term, and $\Delta_{\text{ex}}$ is the exchange energy.
The first term in Eq.\ (\ref{eq:Etot_laan}) originates from spin-conserved virtual excitation in perturbation theory, and the second term from spin-flip virtual excitation. 
Bruno has derived a simplified relation between the MAE and the anisotropy of the orbital magnetic moment by ignoring the second term of Eq.\ (\ref{eq:Etot_laan})\ \cite{Bruno_MAE_3dmonolayer}:
\begin{eqnarray}
 \Delta E 
   &=&\frac {\xi}{4\mu_B}(m^{\text{easy}}_{\text{orb}} - m^{\text{hard}}_{\text{orb}}).
\label{eq:Etot_Bruno}
\end{eqnarray}
Although the Bruno relationship [Eq. (\ref{eq:Etot_Bruno})] has been widely used, it may not be justified to ignore the contribution of spin-flip term in the systems which contain elements with strong spin-orbit coupling such as Pt.
Indeed, recent first-principles calculations on $L{\rm1_0}$-ordered FePt by Solovyev $et\ al.$\ \cite{Solovyev_RSGreens_FePt} and Ueda $et\ al.$\ \cite{Ueda_spinflip_HXPES_FePt} have shown that there is considerable contribution of Pt atoms as well as contribution of the spin-flip excitations to the MAE. The calculations have shown that  the Pt sites contribute to the MAE through hybridization with the Fe 3$d$ orbitals\ \cite{Solovyev_RSGreens_FePt}.

Table\ \ref{tab:table_XMCDvsThesis} summarizes the spin and orbital magnetic moments deduced from the present XMCD experiment and from the first-principles calculation by Solovyev $et\ al.$\ \cite{Solovyev_RSGreens_FePt} using the real-space Green's function method.
The calculated signs of $\Delta m_{\text{orb}}$, positive for Fe and negative for Pt, agrees with the XMCD result, indicating that the first-principles calculation correctly predicts the magnetic properties of $L1_0$-ordered FePt.  From the calculated MAE values decomposed into the Fe and Pt sites (the bottom entry of Table\ \ref{tab:table_XMCDvsThesis}), Solovyev $et\ al.$ concluded that the MAE mainly originates from Pt. The relationship between the anisotropy of the Fe 3$d$ orbital moment ($\Delta m_{\text{orb}} > 0$) and the Fe 3$d$ component of the MAE ($< 0$: in-plane easy magnetization axis) is opposite to that predicted by the Bruno model. As for the Pt site, too, the anisotropy of the Pt 5$d$ orbital moment ($\Delta m_{\text{orb}} < 0$) and the Pt 5$d$ component of the MAE ($> 0$: perpendicular easy magnetization axis) is opposite to the Bruno model. Therefore, the observed anisotropy of the orbital moment indicates that the Bruno relationship\ \cite{Bruno_MAE_3dmonolayer} does not hold even qualitatively for Pt.
 This means that the spin-flip term of Eq.\ (\ref{eq:Etot_laan}), which represents the magnetic dipole term arising from the anisotropic distribution of spin density and is ignored in the Bruno model, should be important to discuss the origin of the MAE of $L1_0$-ordered FePt. The Bruno model has been derived under the assumptions that the spin-orbit coupling is a weak perturbation compared to the exchange splitting and therefore that spin-flip term can be neglected. The importance of the spin-flip term for FePt arises from the large spin-orbit interaction and the small spin splitting of the Pt 5$d$ electrons.
Finally, as for the dependence of the spin and the anisotropy of orbital magnetic moments on the $S$, that is, the observation that $m^{\text{eff}}_{\text{spin}}$ decreases (increases) and $\Delta m_{\text{orb}}$ increases (decreases) for Fe (Pt) sites, respectively, as functions of $S$,
 is similar to the one deduced from the coherent-potential-approximation calculations by Staunron $et\ al.$\ \cite{FePt_CPA_spin} and Kota $et\ al.$\ \cite{CPA_orbitalmoment_FePt}.
%
%
%
%
%
%
\begin{table*}[!t]
\caption{
Spin and orbital magnetic moments deduced from the present XMCD study [corrected for the insufficient saturation for the in-plane magnetization so that $m^{\text{eff}}_{\text{spin}}$ does not depend on the magnetic field direction] and from the first-principles calculation  by Solovyev $et\ al.$ \cite{Solovyev_RSGreens_FePt}. The magnetic anisotropy energy (MAE) of the entire system deduced from magnetization measurements (see supplementary information) and those of individual elements deduced from the first-principles calculation are also indicated. }
\label{tab:table_XMCDvsThesis}
\vspace*{0.3cm}
\begin{ruledtabular}
\begin{tabular}{  c c  c  c  c  c}
\multicolumn{2}{l}{}&\multicolumn{2}{c}{Fe atom}&\multicolumn{2}{c}{Pt atom} \\ 
\multicolumn{2}{l}{}&$m^{\text{eff}}_{\text{spin}}$&$m_{\text{orb}}$&$m^{\text{eff}}_{\text{spin}}$&$m_{\text{orb}}$		\\ \hline
 \multirow{4}{*}{\shortstack{Experiments\\ ($S {\rm= 0.7}$) }}
 			&	 out-of-plane magnetic moment	 & 	\multirow{2}{*}{2.28}		& 	0.21	 	&	\multirow{2}{*}{0.30} 	&	 0.036	\\
			&	 in-plane magnetic moment	 & 		& 	0.12	 	&	 	 	&	 0.043	\\
			&$\Delta m_{\text{orb}} (= m^{\text{out}}_{\text{orb}} - m^{\text{in}}_{\text{orb}})$	
										& 	-- 		& 	0.11	 	&	 --	 	&	 $-$0.005	\\
			 &MAE $\rm{(MJ/m^3)}$
			 							&	\multicolumn{4}{c}{4.6}	\\
 			\hline
			\multicolumn{2}{l}{}&$m_{\text{spin}}$&$m_{\text{orb}}$&$m_{\text{spin}}$&$m_{\text{orb}}$ \\
			\hline
 \multirow{5}{*}{\shortstack{First-principles calculation\\ ($S {\rm= 1}$) }} 			&	out-of-plane magnetic moment	 & 	\multirow{2}{*}{2.77} 		& 	0.0802	 	&	\multirow{2}{*}{0.35}	&	 0.0486	\\
 
			&	 in-plane magnetic moment	 & 	 		& 	0.0690	 	&	 	 	&	 0.0616	\\

			&$\Delta m_{\text{orb}} (= m^{\text{out}}_{\text{orb}} - m^{\text{in}}_{\text{orb}})$	
										& 	-- 		& 	0.0112	 	&	 --	 	&	 $-$0.0130	\\
 	&MAE $\rm{(MJ/m^3)}$
			 							&	\multicolumn{2}{c}{$-$0.887}	&	\multicolumn{2}{c}{14.28}\\
\end{tabular}
\end{ruledtabular}
\end{table*}

In summary, we have performed XMCD studies of $L1_0$-ordered FePt thin films with various $S$ in order to reveal the relationship between the anisotropy of the orbital magnetic moments of Fe and Pt and the magnetic anisotropy. Fe $L_{2,3}$-edge XMCD studies have shown the existence of large anisotropy of the orbital magnetic moment, that is, the orbital magnetic moment is larger for magnetic fields perpendicular to  the plane, which is enhanced with increasing $S$. Pt $L_{2,3}$-edge XMCD studies show that the anisotropy of the orbital magnetic moment is opposite to that of Fe, that is, the orbital magnetic moment is larger for magnetic fields parallel   to the plane than perpendicular to it. This result is qualitatively consistent with the anisotropy of the calculated orbital magnetic moment from first principles  by Solovyev $et\ al.$\ \cite{Solovyev_RSGreens_FePt}, which calculation indicates the significant contribution of Pt and negligible (or even opposite) contribution of Fe to the MAE of the entire FePt film. This is attributed to the strong spin-orbit coupling and small spin splitting of the Pt 5$d$ bands in $L1_0$-ordered FePt. The present result demonstrates that theoretical treatment including spin-flip processes is necessary when designing magnetic materials with large MAE utilizing heavy elements.

\section*{supplementary material}
See supplementary materials for the magnetization curves of each sample.
\begin{acknowledgments}
This work was supported by Grants-in-Aid for Scientific Research from JSPS (grant Nos.\ 15H02109 and 15K17696) and Nanotechnology Platform (project No.\ 12024046) from MEXT. The experiment was performed at BL-16A of KEK-PF with the approval of the Photon Factory Program Advisory Committee (proposal Nos.\ 2013S2-004, 2014G-177, 2016G066, and 2016S2-005) and at BL39XU of SPring-8 with the approval of the Japan Synchrotron Radiation Research Institute (JASRI) (proposal Nos.\ 2014A1158 and 2015B1461). A. F. is an adjunct member of Center for Spintronics Research Network (CSRN), the University of Tokyo, under Spintronics Research Network of Japan (Spin-RNJ). T. S. and K. T. are members of CSRN, Tohoku University, under Spin-RNJ. We thank V. K. Verma and T. Harano for technical support.
\end{acknowledgments}
\bibliography{ref}

\end{document}